\documentclass[dvips]{arxstspdf}
\usepackage{graphicx}
\usepackage{flushend}
\usepackage{stfloats}

\volume{23} \issue{3} \pubyear{2008} \firstpage{321} \lastpage{324}
\doi{10.1214/08-STS244A}



\begin{document}
\begin{frontmatter}
 \vspace{3pt}
\title{Comment: Quantifying the Fraction of Missing Information for
Hypothesis Testing in Statistical and Genetic Studies}
\runtitle{Comment} \referstodoi{10.1214/07-STS244}

\begin{aug}
\author[a]{\fnms{Tian} \snm{Zheng}\ead[label=e1]{tzheng@stat.columbia.edu}}
\and
\author[a]{\fnms{Shaw-Hwa} \snm{Lo}\ead[label=e2]{slo@stat.columbia.edu}}
\runauthor{T. Zheng and S.-H. Lo}

\affiliation{Columbia University}

\address[a]{Department of Statistics, \textit{Columbia University}, New York, New York, USA
\printead{e1,e2}.}

\end{aug}



\end{frontmatter}

\section*{Introduction}

The authors suggest an interesting way to measure the fraction of
missing information in the context of hypothesis testing. The
measure seeks to quantify the impact of missing observations on the
test between two hypotheses. The amount of impact can be useful
information for applied research. An example is, in genetics, where
multiple tests of the same sort are performed on different variables
with different missing rates, and follow-up studies may be designed
to resolve missing values in selected variables.

In this discussion, we offer our prospective views on the use of
relative information in a follow-up study. For studies where the
impact of missing observations varies greatly across different
variables and where the investigators have the flexibility of
designing studies that can have different efforts on variables, an
optimal design may be derived using relative information measures to
improve the cost-effectiveness of the follow-up.

Using the simple motivation example in their paper, we examine the
estimation of relative information by $\mathcal{R}I_1$ and $\mathcal
{R}I_0$ in
terms of unbiasedness and variability, and discuss issues that
require further research. Although the relative information measure
developed in their paper estimates the mean impact of the missing
data, the actual impact may be highly variable when the amount of
information in the observed data is moderate or small, which makes
the estimated mean relative information a less reliable prediction
of the actual impact of missing observations. For this reason, we
suggest a simple way to estimate the variability of relative
information between complete data and observed data in the simple
motivation example. Further investigation is required in
incorporating these variability estimates into the optimal design of
follow-up studies.

\section*{Relative Information and Follow-up Study Designs}

Missing values can occur for many reasons and can have different
effects on a given test. Nicolae, Meng and Kong pointed out that the
impact of missing values (in terms of {\em relative information}) on a
test may not be as simple as the ``face value'' of $n_0/n$, where $n_0$
is the number of observed values and $n$ is the number of individuals
($n-n_0$ is then the number of missing values). Therefore, a more
accurate estimation of the information gain due to the resolution of
missing values is important for the design of follow-up studies.

Given an existing data with $n$ individuals (with missing values),
if $n_1$ additional independent samples are collected (possibly with
the same missing rate) to expand this data set, it is intuitive to
assume that the ratio of information in the original data and the
expanded data is approximately $n/(n+n_1)$. Now consider a test on
the existing data with $n$ individuals that has some missing values
(say, $n_0$ observed values). The {\em relative information} is
estimated to be 80\%, meaning that if the data used for this test is
``resolved'' to become complete, the expected log likelihood ratio
is about $1/80\%=125\%$ of the observed log likelihood ratio. To
achieve the same level of information by adding new independent
observations, one would need to collect a sample of additional
$n_1=n \times25\%$ individuals. In many situations, resolving
missing values, if possible, turns out to be much cheaper than
collecting data on additional samples. In Section 2 of the NMK
paper, an example was given on genotyping ambiguity in genetic
linkage analysis (meaning that the exact inheritance vectors needed
for the lod score computation cannot always be derived given the
genotypes observed on the individuals). Here, let $Y_{\mathrm{ob}}$ be
current data with unambiguous genotypes. For a follow-up study, a
researcher can decide between (1) increasing the density of genetic
markers on the observed individuals to {\em resolve} the
ambiguities and (2) increasing the sample size by genotyping more
independent individuals on the same set of markers for the
previously observed individuals. If we denote the two potential
expanded data sets as $Y_{\mathrm{co},m}$ and $Y_{\mathrm{co},i}$
with $m$ and $i$
standing for {\em markers} and {\em individuals}, we can compute the
fraction of information between $Y_{\mathrm{ob}}$ and $Y_{\mathrm
{co},m}$, and
between $Y_{\mathrm{ob}}$ and $Y_{\mathrm{co},i}$, potentially using
$\mathcal{R}I_1$ and
$\mathcal{R}I_0$ proposed in the NMK paper. Comparing these two
measures of
relative information, the researcher can then decide which option
(increasing markers or increasing individuals) is cost-efficient for
the inferential task at hand.

In practice, one would need to consider such comparison at multiple
variables simultaneously. Here we consider a simple example. Let
$\{Y_1, \ldots, Y_M\}$ be the variables studied. For $Y_i$,
$n_{0,i}$ values are observed on $n$ individuals. In a follow-up
study $n_{1,i}$ missing values can be resolved at $Y_i$. At $Y_i$,
the relative information (say, $\mathcal{R}I_1$) is a function of $n_{1,i}$,
the observed $\mathrm{lod}$ score $\mathrm{lod}_{\mathrm{ob},i}$
and the observed m.l.e. To
evaluate the {\em overall} information gain due to these additional
observations, we suggest an expression similar to that of
(19) in the NMK paper\footnote{Equation (19) in the original paper
is to combine relative information measures from several studies,
while  (\ref{eqn:optdesign}) here is to evaluate relative
{\em overall} information of multiple variables.}:
\begin{eqnarray}\label{eqn:optdesign}
&&\overline{\mathcal{R}I_1}^{-1}(n_{1,1}, \ldots, n_{1,M})\nonumber
\\[-8pt]\\[-8pt]
&&\quad=\frac{\sum _{i=1}^M \mathrm{lod}_{\mathrm{ob},i}\mathcal
{R}I_1(n_{1,i})^{-1}}{\sum_{i=1}^M
\mathrm{lod}_{\mathrm{ob},i}}.\nonumber
\end{eqnarray}
A possible way to yield an optimal design would be to select values
of $0 \le n_{1,i} \le n-n_{0,i}$ to maximize the information gain
while controlling for a fixed cost. Differences in design may
involve varying setup costs that may depend on, for example, the
number of nonzero $n_{1,i}$ such as that in genotyping studies.
Once such a cost function can be fully specified, linear programming
can be used to obtain the optimal design. If the $n_{1,i}$'s in the
optimal design identified take similar values on $i=1,\ldots,M$,
this may suggest a design that collects data on $n_1$ new
independent individuals and takes measurements on the same $M$
variables as in the original data.

Another advantage of the likelihood ratio-based evaluation of
information used by Nicolae, Meng and Kong is that one can evaluate
the potential information gain conditioning not only on the observed
data at the current concerned variable but also on some associated
variables, through a model-based calculation. Similar model-based
strategies have been commonly used for imputing missing genotypes in
genetic studies. Such consideration may introduce more complicated
design questions than the computation in (\ref{eqn:optdesign}) but
may also bring better efficiency.

\section*{The ``Empirical'' Fraction of Information and Its Variability}

Using the simple motivation example in Section 1 of the NMK paper,
we consider the relation between the empirical observed data log
likelihood ratio (lod score) and the ``random'' complete data log
likelihood ratio (lod score). We offer relationships between the
proposed fraction of information and the distribution of the
``empirical'' ratio. The ``empirical'' ratio is the actual random
gain due to additional observations, while the estimation of
relative information and the possible optimal design derived are
intended to approximate this random outcome.

\begin{figure*}[t]

\includegraphics{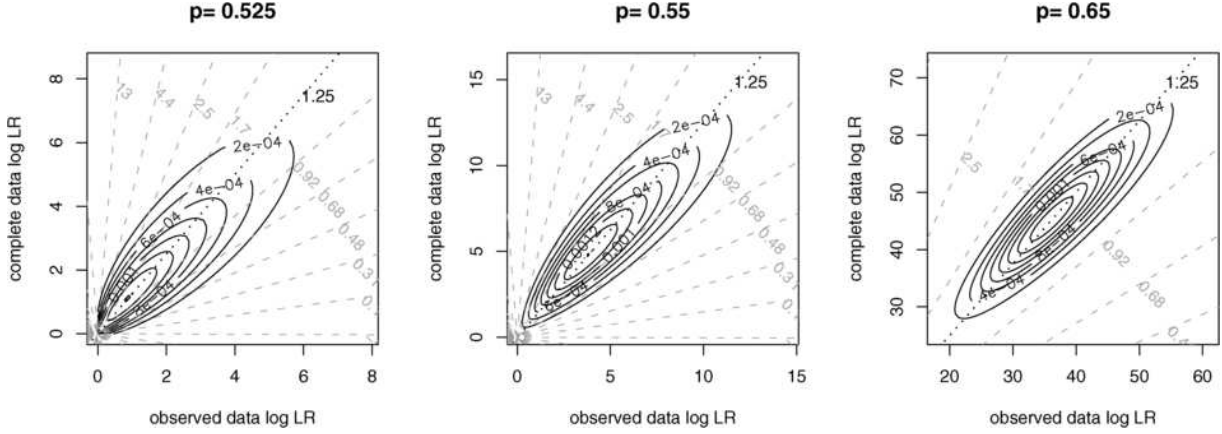}

\caption{Distribution of log likelihood ratio
test statistics (or lod scores) given observed data and complete
data. The contour plots display the joint distribution of the log
likelihood ratio test statistics given the observed data and the
complete data. Given $n_0=800$ and $n=1000$, the ratio between the
complete data log LR and the observed data log LR is expected to be
$n/n_0=1.25$. In each contour plot, a dotted line is plotted to
indicate the $y=1.25x$ line. The gray broken lines display $y=rx$
with $r$ varying and provide reference for the empirical ratio of
the complete data log LR and the observed data log LR. }
\label{fig:contour}
\vspace*{12pt}
\end{figure*}

In Figure~\ref{fig:contour}, we plot the joint distribution of the lod
scores under the observed data and the complete data, with missing
percentage being 80\%. The distribution is evaluated under three true
values of the probability of success with $n_{0}=800$ and $n=1000$. To
obtain a realistic evaluation, we use the traditional definition of the
likelihood ratio test (or the lod score) where the ratio is evaluated
between the maximum likelihood estimate given {\em current} data
(observed or complete) and the value in the null hypothesis.

We first notice the positive correlation between the complete data
statistic and the observed data statistic. Gray broken lines in
Figure~\ref{fig:contour} give reference lines for empirical or
``random'' ratio between the complete data lod score (or log LR
statistic) and observed lod score. The estimated $\mathcal{R}I_1$ (which
coincides with $r=n_{0}/n$) corresponds to a line going through the
center of the joint distribution (almost exactly), indicating it is
a good estimate for the expected ratio (or fraction of information)
regardless of the values of the observed lod score.

For a small departure (say, $p=0.55$) from the null hypothesis
($p_0=0.5$), the LR test does not have great power and the test
statistics distribute close to zero. The contour of the distribution
intersects with lines whose ratio values are shown to go as high as 13.
This is natural given the observed data statistic can become very small
due to chance and create a highly variable ratio. For values that are
far away from the null hypothesis, the estimated $\mathcal{R}I_1$
becomes more precise.

\begin{figure*}[t]
\includegraphics{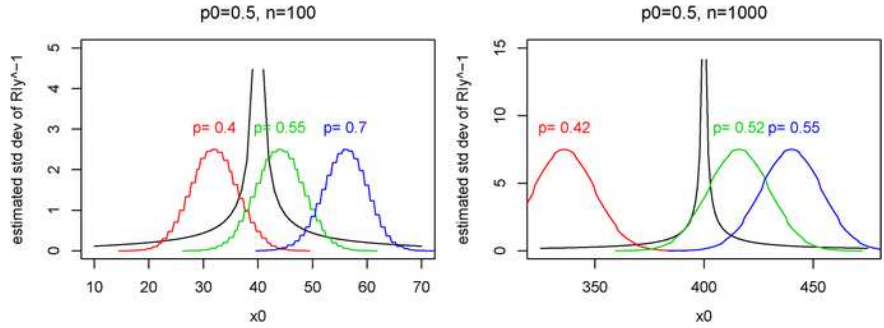}
 \caption{Estimated standard deviation of
$\mathcal{R}I_{y}^{-1}$. For sample size $n=100, 1000$, we plot the
estimated standard deviation of $\mathcal{R}I_{y}^{-1}$ against the
observed number of successes $x_0$. Density curves of observed number
of successes $x_0$ under different true $p$ values are plotted.}
\label{fig:rivar}
\end{figure*}

As illustrated above and in Figure~\ref{fig:contour}, the unobserved
random missing values make the relative ``empirical'' information a
random quantity. It is instructive to evaluate the amount
of\vadjust{\goodbreak}
variation in the complete data lod score. It is easy to obtain for
the simple binomial example that
\begin{eqnarray}\label{eqn_generalV}
&&\mathrm{var}[\mathrm{lod}(p_1, p_2; Y_{\mathrm {co}})|
Y_{\mathrm{ob}}, p]\nonumber
\\[-8pt]\\[-8pt]
&&\quad=(n-n_{0})p(1-p)\biggl[\log\frac{p_{1}}{p_{2}} -\log\frac
{1-p_{1}}{1-p_{2}}\biggr]^{2}.\nonumber
\end{eqnarray}

Consider a null hypothesis that specifies the probability of success
as $p_{0}$ and let $p$ be the true parameter value. Let
$\mathcal{R}I_{y}(Y_{\mathrm{co}}, Y_{\mathrm{ob}};p,
p_{0})$ be the empirical
fraction of information regarding the difference between $p$ and
$p_{0}$, for a set of $Y_{\mathrm{co}}$ with only $Y_{\mathrm{ob}}$
observed ({\em
or} the ratio of the lod scores between $p$ and $p_{0}$ derived
using the observed data and the {\em potential} complete data). It
is easy to see that $\mathcal{R}I_{y}^{-1}$ is a more natural relative
information ratio to use for evaluating overall relative information
in (\ref{eqn:optdesign}) and identifying optimal follow-up
design. From similar computation in (\ref{eqn_generalV}),
$\mathcal{R}I_{y}^{-1}$, conditioning on $Y_{\mathrm{ob}}$, has an expectation
\begin{eqnarray*}
&&\mathrm{E}\mathcal{R}I_{y}^{-1}
\\
&&\quad=1+(n-n_{0})\biggl[p\log \frac{p}{p_{0}}+(1-p)\log
\frac{(1-p)}{(1-p_{0})}\biggr]
\\
&&\qquad\ \,\quad{}\cdot{(\mathrm{lod}(p, p_0; Y_{\mathrm {ob}}))^{-1}}
\end{eqnarray*}
and variance
\begin{eqnarray*}
\mathrm{var}\,\mathcal{R}I_{y}^{-1}&=&(n-n_{0})p(1-p)\biggl[\log
\frac{p}{p_{0}} -\log\frac{1-p}{1-p_{0}}\biggr]^{2}
\\
&&{}\cdot{(\mathrm{lod}\left(p, p_0; Y_{\mathrm{ob}
}\right)^{2})^{-1}}.
\end{eqnarray*}
In practice, we may substitute $p$ with $\hat{p}_{\mathrm{ob}}$
and have $\widehat{\mathrm{E}\mathcal{R}I_{y}^{-1}}$
estimated by $\mathcal {R}I_1^{-1}$.
Figure~\ref{fig:rivar} gives the estimated standard deviation of
$\mathcal{R}I_{y}^{-1}$ with probability density curves under different
true values of $p$. When the true value is close to the null hypothesis
$p_0$, $\mathcal{R}I_{y}^{-1}$ is highly variable, which will make the
simple estimate of $\mathcal{R}I_1^{-1}$ as an estimated expectation of
$\mathcal{R}I_{y}^{-1}$ a unreliable prediction of
$\mathcal{R}I_{y}^{-1}$. A procedure incorporating both
$\widehat{\mathrm{E}\mathcal{R}I_{y}^{-1}}=\mathcal {R}I_1^{-1}$ and an
estimated standard error of $\mathcal{R}I_{y}^{-1}$ should be
considered to address the design issues similar to that of
(\ref{eqn:optdesign}).

\section*{In Summary}

The paper by Nicolae, Meng and Kong provides interesting evaluation
strategies for relative information discerning two hypotheses contained
in observed data. Such measures support the quantification of possible
information gain that can be brought by additional observations, which
can be used to optimally design follow-up efforts. The measures
$\mathcal{R}I_1$ and $\mathcal{R}I_0$ deserve more research for further
understanding. More importantly, theory and practice\break should be
incorporated to provide design suggestions that\vadjust{\goodbreak}
utilize relative information such as $\mathcal{R}I_1$ and corresponding
variability measures.

\section*{Acknowledgments}
This research is supported by NIH Grant R01 GM- 070789 and NSF Grant
DMS-07-14669.

\end{document}